\documentclass[english,preprint,endfloats,floatfix,amsmath,amssymb,aps]{revtex4}
\usepackage{subfigure}
\usepackage{graphics}

\usepackage{graphicx}
\usepackage[T1]{fontenc}
\usepackage[latin9]{inputenc}
\usepackage{ifsym}
\usepackage{wasysym}
\usepackage{babel}
\usepackage{color}

\begin{document}

\title{Precise measurements of torque in von Karman swirling flow driven by a bladed disk}

\author{Aryesh Mukherjee$^a$, Sergei Lukaschuk$^b$, Yuri Burnishev$^a$, Gregory Falkovich$^a$, and Victor Steinberg$^{a,c}$}
\affiliation{$^a$Department of Physics of Complex Systems, Weizmann Institute of Science, Rehovot 76100, Israel,\\
$^b$University of Hull, Hull, UK,\\
$^c$The Racah Institute of Physics, the Hebrew University of Jerusalem, Jerusalem 91904, Israel}

\date{\today}

\begin{abstract}
Scrupulous measurements and detailed data analysis of the torque in a swirling turbulent flow driven by counter-rotating bladed disks reveals an apparent breaking of the law of similarity. Potentially, such breakdown could arise from several possible factors, including dependence on dimensionless numbers other that $Re$ or velocity coupling to other fields such as temperature. However, careful redesign and calibration of the experiment showed that this unexpected result was due to background errors caused by  minute misalignments which lead to a noisy and irreproducible torque signal at low rotation speeds and prevented correct background subtraction normally ascribed to frictional losses.  An important lesson to be learnt is that multiple minute misalignments can nonlinearly couple to the torque signal and provide a dc offset that cannot be removed by averaging. That offset can cause the observed divergence of the friction coefficient $C_f$ from its constant value observed in the turbulent regime. To minimize the friction and misalignments, we significantly modified the experimental setup and carried out the experiment with one bladed disk where the disk, torque meter and motor shaft axes can be aligned with significantly smaller error, close to the torque meter resolution. As a result we made precise measurements with high resolution and sensitivity of the small torques produced for low rotation speeds for several water-glycerin solutions of different viscosities and confirmed the similarity law in a  wide range of $Re$ in particular in low viscosity fluids.

\end{abstract}

\maketitle

\section{Introduction}

\indent The von Karman swirling flow between two rotating coaxial impellers has been extensively studied over the last two decades \cite{fauve,tabeling,cadot,labbe,pinton}. In this flow, disks of radius $R$ counter-rotating with the same angular velocity $\Omega$ (rad/s)  are used to mechanically stir the fluid, thereby injecting energy in the bulk from the surface of the disks. Experiments have shown that attaching blades to the disk surface are more effective at injecting energy, compared to a smooth surface (known as viscous stirring), and the energy is directly supplied to the vortices trailing behind the blades and with a size comparable to the disk. Known as inertial stirring, such method of energy injection can create very high turbulence levels with velocity fluctuations up to 50\% of the blade velocity \cite{ravelet}. The blades can be flat or curved, where in the latter case the different direction of curvature of the blades relative to the rotation can lead to different dynamics \cite{ravelet}. Although such a flow is far from the theoretically explored regime of homogenous, isotropic turbulence in an infinite space, it has been experimentally shown to reproduce the theoretically predicted scaling relation of injected and dissipated energy \cite{tabeling,ravelet,burnishev}.  Besides, inertial  stirring also sets a well defined spatial scale for energy injection, almost an order of magnitude larger than that for viscous stirring \cite{burnishev,burnishev1}.

\indent  A closed rotating flow, as described above, and its transition to turbulence can be quantitatively characterized by torque measurements. In such flows, the normalized average torque $\bar{\Gamma}$,  the friction coefficient, $C_f=\bar{\Gamma}/\rho\Omega^2 R^5$, where $\rho$ is the fluid density should solely be a function of $Re=\rho\Omega R^2/\eta$ ($\eta$ is the fluid dynamic viscosity) according to predictions from the law of self similarity.  Earlier torque measurements performed in very high viscosity water-glycerin solutions \cite{ravelet}with glycerin concentrations from $74\%$ up to $99\%$ w/w and two temperatures $15^\circ$C and $30^\circ$C, show that $C_f$ decreases from $Re\simeq50$ till $Re_c\simeq3.3\times 10^3$, where a transition from laminar to a fully developed turbulent regime takes place and becomes constant valued, independent of $Re$. However, the data for $C_f$ of water-glycerin solutions in a laminar as well as transition region overlap within  large error bars and do not provide the requisite accuracy for such measurements. Moreover, no data on lower viscosity solutions were reported, except for a single point in the turbulent regime in water at $Re>10^6$ \cite{ravelet,cadot}.

\indent More detailed data on $C_f$ in the turbulent regime taken solely for water-sugar solutions at different sugar concentrations from $0\%$ up to $40\%$ at a constant temperature of $24\pm0.1^{\circ}$C corresponding to the low dynamic viscosities $\eta$ from about $1$ up to $6.1~mPa\cdot s$ were presented in Ref. \cite{burnishev}, where only the constant values of $C_f$ in a wide range of $Re$ are shown.

\indent Contrary to the data reviewed above, torque measurements in water-sugar solutions  at constant temperature conditions by a torque meter of much higher resolution and accuracy than those used in all former experiments showed a surprising deviation of $C_f$ from the constant value for each solution at low $\eta$ and $\Omega$  and corresponding value of $Re$, supposedly in the fully turbulent regime and depended on the solution viscosity $\eta$ \cite{unpublished_data}. Hence the data shows that $Re_c$ at the laminar-to-turbulent transition for solutions of different sugar concentrations depends explicitly on the solution viscosity meaning that $C_f$ changes due to viscosity variations even if the value of $Re$ is kept invariant. Indeed, in the range of $Re$ between $\sim 5\times 10^4$ and $10^6$ several values of $C_f$ for a single value of $Re$ were found \cite{unpublished_data}. This is an apparent violation of the law of similarity according to which $C_f$ must depend on $Re$ only.

\indent In the 19 century, experimental measurements of the pipe flow resistance by Reynolds inspired Stokes and Rayleigh to suggest that similarity must hold for the mean resistance in the case of laminar as well as unsteady and turbulent flows. The same conclusion {had been also} reached earlier by Helmholtz \cite{Dar}. Measurements in numerous flows supported the law of similarity, which  is {now} the foundation of modeling in engineering - one builds a small model of ship or plane and  {places} it  {in a} wind tunnel or basin {respectively} to measure resistance. This 19-century law presented in every textbook on fluid mechanics \cite{batchelor,landau,falkovich} is of immense practical importance in engineering (allowing for re-scaled measurements) and deeply influenced modern physics, which is now permeated by similarity and dimensional reasoning.

\indent Several reasons can be given for the observed blatant violation of the fundamental law of similarity. First, data for the torque {may not have been} collected for {a} sufficiently long time and thus not averaged for a sufficiently long enough time scale to assure asymptotically stationary statistics such that the mean torque is independent of the averaging time. Second, the similarity law can also be broken if the fluid velocity field interacts with another field, the most obvious of which is a temperature field. In the turbulent state giant velocity fluctuations can lead to a non-uniform temperature distribution due to non-uniform dissipation and generate heat fluxes. Thus it is possible that $Re$ is not the only dimensionless control parameter in the system, and that others like the Prandtl number $Pr=\eta/\rho\kappa$  and the Rayleigh number $Ra=\Delta T l^3 g\alpha\rho/\eta\kappa=\Delta T l^3 g\alpha\rho^2 Pr/\eta^2$ couple the velocity of a fluid flow to heat fluxes causing the apparent violation of self-similarity. Here $\alpha$ and $\kappa$ are fluid thermal diffusivity and thermal expansion coefficients, respectively, $g$ is the gravitational constant, and $\Delta T$ is the temperature difference on the spatial scale $l$. A third possible reason is the low resolution of the torque measurements due to large irreproducible and unpredictable background errors at the level of the torque signal in spite of the fact that our torque measurements are more precise than those reported in all previous experiments in a swirling flow (see \cite{burnishev}). Besides our recent measurements in a  swirling flow  driven viscously by a smooth disk showed the expected self-similarity, i.e. $C_f$ depends on $Re$ solely in a wide range of water-glycerin solution viscosities \cite{aryesh}. Thus the main goal of these studies is minimize background errors below that of the small torque signals thereby improving upon previously achieved signal to noise ratios at very small signal values and measure the friction coefficient as a function of $Re$ with accuracy of the torque meter resolution.

\indent The paper is organized as follows. We first present the measurements of torque, pressure fluctuations, and flow velocity in the swirling flow of water-glycerin solutions in a wide range of viscosities driven by two counter-rotating, curved bladed disks in the experimental setup similar to that already reported in Ref. \cite{burnishev}. In these measurements a sequence of laminar-to turbulent transitions were observed with corresponding critical $Re$ depending on the fluid viscosity. We also describe pressure fluctuations and local two-component velocity measurements simultaneously carried out in a wide range of $Re$ which do not provide any evidence of the laminar-turbulent transition, except {for the one known to occur} at $Re\approx3.3\times 10^3$.  Then we report the detailed precise torque measurements with even a more sensitive torque meter first in a swirling flow driven by two-counter-rotating bladed disks at low $\Omega$ which are followed by the precise torque measurements in a swirling flow driven by a single bladed disk of the same solutions in a significantly modified setup. The latter is designed to perform a precise alignment between the upper disk, the torque meter, and the motor axes and to conduct the torque measurements with almost two orders of magnitude higher resolution and accuracy than in our previous experiments. A detailed discussion of background errors due to friction losses but mostly due to misalignments is provided with estimates and experimental proof that the fine alignment leads to the expected self-similarity within the torque meter resolution.

\section{Experimental setup}

\indent In these studies we used two different setups.

\indent The first setup is the same as used in previous experiments and reported in Ref. \cite{burnishev}. It consists of  a cylindrical vessel where {the} swirling flow is generated made of Lucite (Plexiglas) of diameter $D=29$ cm and height $L=31$ cm surrounded by a hexagon shaped Lucite enclosure with a gap between them filled with water for thermal regulation. Two 10 mm thick Lucite disks of $R=13.2$ cm and $H=19$ cm apart were used as impellers and had 8 curved blades of 10 mm height attached to it with radius of curvature 7 cm corresponding to a blade inclination angle of $\alpha=70.5^{\circ}$ (see Fig. \ref{fig:setup1}). The stirrers rotate with the concave face of the blades pushing the fluid driven by two brushless sinusoidal motors F-6100 (Electro-Craft servo systems) with maximum continuous torque of $13$ Nm and peak torque $31.1$ Nm controlled  by a motion card PCI-7344NI (Advanced Motion Controls) via optical encoders with constant velocity within $0.1\%$ up to $\Omega/2\pi=9$ Hz ($540$ rpm) limited by the overheating of the motion control card. Two different seals were used to prevent water leakage from penetrating into the ball-bearings: a rotating SS-R00 John crane type R00, AES P08, wave spring design, balanced end-face mechanical seal with guarantied limited mechanical friction at the bottom motor driving shaft and an oil seal CR14223 with low mechanical friction from CR Services Co at the top motor driving shaft.

\indent The measured average torque $\bar{\Gamma}_0$ was corrected for background errors that consists of friction losses and other {systematic} sources of errors $\bar{\Gamma}_{err}$ obtained from the torque measurements in air {with} disks with  curved blades yielding the following expression $\bar{\Gamma}=\bar{\Gamma}_0-\bar{\Gamma}_{err}$. Estimates show that the moment of inertia of water is about 50 times larger than the total moment of inertia of the rotor, the bladed stirrer, and the torque meter resulting in a cut-off frequency of the control of about $20$ Hz. Thus, in constant angular velocity mode with the feedback loop control the torque spectral measurements are reliable up to about $10$ Hz. The temperature of the fluid was measured by three thermistors placed at three different heights along the side of the container, at 2cm, 20cm and 28cm respectively measured from the top lid using a RTD thermometer for calibration. The temperature was  stabilized within $\pm 0.1^{\circ} C$ by three circulating refrigerators (Lauda Inc.) driving cooling water via coiled copper tubes soldered to the top and bottom cylinder lids made of stainless steel and via the space between the hexagon shaped Lucite enclosure and the cylindrical vessel.

\indent  The torque required to generate the swirling flow was measured using a torque meter inserted between the upper disk and the motor (see Fig. \ref{fig:setup1}) when the  disks rotate in opposite directions at equal rates, by a non-contact calibrated torque meter MCRT 49001 (S. Himmelstein Co.) within the range up to 11.3 Nm (with overload up to 45.2 Nm) with 0.1\% accuracy and up to 15000 rpm. The torque meter has two low-pass filter outputs: one with bandwidth between dc and 500Hz, which we used in the measurements, and another one with bandwidth between dc and 1 Hz used to verify the average value of torque {with} output rms noise  $0.1\%$ and $0.01\%$ of the full scale respectively. For a second, more precise measurement using the same setup, another torque meter MCRT no. 48999VB(5-1)NFNN (S. Himmelstein Co.) with a range up to 0.35 Nm and overload up to 0.7 Nm with $0.1\%$ resolution was used. The torque was measured at constant rotation rates of the disks $\Omega$ stabilized through feedback control and optical encoder and used to test the stability of the measurements via deviation of ${\Gamma}$ from constant value due to temperature variations or other factors.

\indent The pressure fluctuations  were measured by a 40PC001B miniature signal conditioned pressure sensor (Honeywell Co.) with 2.8 mm diameter membrane and working pressure range up to $\pm 6.7$ kPa with accuracy up to 0.2\%, resolution 0.3 mV/Pa and rise time less than 1 msec. However, the limited working pressure range of this pressure sensor {did} not allow us to use it at sufficiently high $Re$. Both torque and pressure data were acquired with a sampling rate of about $83$ Hz to avoid high-frequency noise, more than 4 times larger than the cut-off frequency mentioned above. Since we were not interested  in studying the frequency power spectra of pressure in an inertial range but just to observe the transition to  turbulence, the frequency range below 100 Hz  was sufficient to meet our goals.

\indent In the first experimental setup three different experiments with water-glycerin solutions of different concentrations, temperatures, and so different dynamic viscosities $\eta$ as working Newtonian fluids were conducted. The values of these parameters of the corresponding water-glycerin solutions are presented in Table 1, Table 2, Table 3. The values of $\eta$ were obtained via our rheological measurements.

\indent Important information on local measurements of the average values $V_x$ and $V_y$ and rms fluctuations $V_x^{rms}$ and $V_y^{rms}$ of two velocity components as a function of $Re$ were obtained by laser Doppler velocimetry (LDV) measurements using a two-component DANTEC LDV system including a two-component DANTEC FiberFlow fiber-based flow probe, transmitter, Coherent Innova 70 Ar-ion laser, and BSA-F30 Flow Processor. Two pairs of beams (488 nm and 514.4 nm) from the flow probe were focused onto a small fluid volume of $0.1\times 0.1\times 2~ mm^3$ seeded with 20$\mu m$ Polyamide Seeding Particles (from DANTEC). The measurements were conducted in a wide range of $Re$ from $\sim150$ and up to $5\times 10^5$ in the water-glycerin solutions which properties are presented in Table 2.

\indent The second experimental setup consisted of the same apparatus to generate a swirling flow, however only an upper bladed disk was used driven by {the same} brushless DC motor as previously used. In this case, the rotating disk was separated from the cylinder bottom considered as a stationary disk by a distance $H=24$ cm ($R/H=0.55$) (see Fig. \ref{fig:setup2}). The torque applied on the upper disk to generate the swirling flow was measured by the same non-contact calibrated, more precise torque meter MCRT no. 48999VB(5-1)NFNN (S. Himmelstein Co.) mentioned above.

\indent In order to conduct precise measurements at low frequencies and extend the range of the torque meter, both the friction and misalignment had to be minimized. High precision machining and low friction ball-bearings helped alleviate the first problem. To minimize the errors of the second kind the following modifications were made. Traditionally soft helical spring couplers have been used as a solution to slight misalignment between two axes. Attaching spring couplers in our system caused the measured torque to oscillate at particular drive frequencies due to unavoidable non linear resonances. For accurate calibration the torque meter rotation axis was precisely aligned to that of the disk shaft by attaching the former to a two axis rotation stage attached to a XY translation stage. This configuration allowed enough degrees of freedom to align both the position and the angle of the torque meter.

\indent The radius of the disk shaft was 12.5 mm while that of the torque meter was 3.15 mm. The coupler from the disk to torque meter was machined out of aluminum and was tapered down, being wide (31mm) at the bottom and narrow at the top (9.5mm). This coupler design lowered its center of mass considerably and reduced nutation errors. At the top of the coupler a narrow notch (width 8 mm) was machined, into which the corresponding protrusion from the torque meter fit in precisely, analogous to a flat head screw driver and screw. This arrangement minimized bending torques between the torque meter and the disk shaft but caused a small mismatch in the shaft angles between the torque meter and the DC motor. The centers of the motor and torque meter were aligned by placing the former on a XY translation stage while errors due to angle misalignment was minimized by connecting the motor and torque meter with a two arm connector which allowed two extra degrees of freedom. After these modifications the background DC signal was stable to within 0.2 mNm, close to the precision allowed by the instrument. We also observed that at a low rotation speed the background torque does indeed increase due to a persistent misalignment owing to a nutation of the disk shaft coupler (this occurs because the shaft is not perfectly aligned with gravity). In the second experimental setup, water-glycerin solutions of different concentrations, the same temperature and so different dynamic viscosities $\eta$ as working Newtonian fluids were used. The values of these parameters are presented in Tables 4.

\section{Results}

\subsection{Torque and pressure scaling relations and velocity measurements in a swirling flow driven by two counter-rotating disks}

\subsubsection{Large {range} torque meter}

\indent For the first data set we obtained average torque $\bar{\Gamma}_0$ as a function of the rotation frequency $f$ ($f=60\Omega/2\pi$) with the large {range} torque meter in a wide range of $f$ from 1 up to 540 rpm and $\eta$ from 0.6 up to $1432~mPa\cdot s$ for water-glycerin solutions from zero up to $99\%$ w/w of glycerin and various temperatures (see Table 1). The data is shown in Fig. \ref{fig:Tvsomegaoldfull}, and has a $\sim f^2$-dependence which implies that the friction factor $C_f$ is independent of $Re$ as expected in a turbulent regime. However, on a smaller scale of $f$ from 0 up to 60 rpm the data shows large scatter and an offset from zero $\bar{\Gamma}_0\leq 0.3~Nm$ at $f\rightarrow 0$ (see inset in Fig. \ref{fig:Tvsomegaoldfull}) which is surprising considering that the torque meter resolution and accuracy is about 30 times less than previously reported measurements. Similar behavior is presented Figs. 1SM and 2SM in \cite{sm}, where analogous plots for water at $T=45^{\circ}$C and water-glycerine solution with $50\%$ w/w glycerine at $T=24^{\circ}$C are shown. Moreover, at $f<30$ rpm in water an abnormal dependence of $\bar{\Gamma}_0$ on $f$ and large scatter stand out in particular (see insets in Fig. 1SM in \cite{sm}).

\indent To further investigate the anomalous $\Gamma$ dynamics at low rotation frequencies, we collected very long time series of $\Gamma$ of a solution with $60\%$ w/w glycerin at $24^{\circ}C$ (Fig. \ref{fig:Tvst60Gl24C}). The data displays rare and strong bursts of $\Gamma$, much larger than the torque meter resolution and may provide evidence of non-stationary $\Gamma$ statistics which is also unexpected  taking into account the observation time of about 120 hours. On the other hand, at higher $Re$ and for a solution contained $20\%$ w/w glycerin at $21^{\circ}$C a time series has much smaller scatter of about $\pm 20$ mNm in $\Gamma$ and a stable average value. Similar time series are shown in Fig. 3SM in \cite{sm} at higher values of $Re$ for the solution contained $20\%$ w/w glycerin at $21^{\circ}$C.

\indent To accurately calculate the friction factor the background errors, $\Gamma_{err}$, have to be subtracted from the average torque $\bar{\Gamma}_0$ shown in Fig. \ref{fig:Tvsomegaoldfull}. Theoretically, background friction errors should not depend on rotation speed, however, we find the measurements of $\Gamma_{err}$ are noisy and irreproducible, in particular at low $f$ which significantly affects $C_f$ at low $f$. Instead of using $\Gamma_{err}$ we shifted down $\bar{\Gamma}_0$ by $0.3~Nm$, which is the offset in average torque as $f\rightarrow 0$. The resulting  $C_f=\bar{\Gamma}/\rho\Omega^2 R^5$ as a function of $Re$ is presented in Fig. \ref{fig:CfvsReold}. Our measurements  are in fair agreement with the results published in Ref. \cite{ravelet} for high viscosity water-glycerin solutions, the transition to turbulence occurs around $Re_c\approx 3.3\times 10^3$, and the turbulent regime is characterized by a constant $C_f$ for any $Re>Re_c$ up to about $Re\approx 10^4$ as expected from the law of similarity (see Fig. 4SM in \cite{sm}). However, for low viscosity water-glycerin solutions our measurements showed anomalous divergence of $C_f$, between at $Re\geq 4\times 10^4$ up to $Re \approx 2\times 10^6$ not only as a function of $Re$ but also as a function of $\eta$, an apparent violation of self-similarity (see Fig. \ref{fig:CfvsReold}).

\indent The next set of experiments were conducted with the same set up but at $\Omega\geq 3$ rad/s (or at $f\geq 28.7$ rpm). The data for $\bar{\Gamma}_0$ {is} smooth with negligible scatter in the plot (Fig. \ref{fig:TvsOmegalarge}) but  still offset from zero as $\Omega\rightarrow 0$, (inset of Fig. \ref{fig:TvsOmegalarge}). As in our early measurements, $\Gamma_{err}$ reaches large values up to $0.3$ Nm and is irreproducible at the level of $\sim50$ mNm, as shown in the inset in Fig. \ref{fig:CfvsRelarge}. Hence, to compute  $C_f$ as a function of $Re$, we shifted the average torque in Fig. \ref{fig:CfvsRelarge} by the zero frequency offset, as in previous case (see Fig. \ref{fig:CfvsReold}), and found that the blatant breaking of the self-similarity persisted. However, in this case the data collapses on a  single curve (see Fig. \ref{fig:CfvsReHlarge}) by rescaling $Re$ by a function $H(\eta/\rho)$ (i.e. the solely function of $\nu=\eta/\rho$) shown in the inset in Fig. \ref{fig:CfvsReHlarge}.  Concurrently, $\Gamma_{rms}/\bar{\Gamma}$  for water-glycerin solutions in a whole range of viscosities show  strong divergence from almost $Re$ independent behavior at each $\eta$ (see Fig. \ref{fig:TrmsTvsRelarge}). To summarize, despite the higher resolution and accuracy of the torque meter used  than all previous measurements and improve the torque measurements resolution by using the data only above some rotation velocity, the observed violation of the similarity law persists.

\indent  However, simultaneous measurements of $p_{rms}$ (in the first run) shown in its scaled form as $C_p=p_{rms}/\rho\Omega^2R^2$ versus $Re$  in Fig. \ref{fig:CpvsReold} and of two components of velocity and rms fluctuations of velocity (in the second run) presented in Fig. \ref{fig:VxVyvsRe} do not provide any evidence of additional transitions in the flow at $Re>3300$. Indeed, in Fig. \ref{fig:CpvsReold} in the plot of $C_p$ versus $Re$ for several solution viscosities $C_p$ is independent of $Re$ from $Re\geq2\times 10^4$ up to $7\times 10^5$, in the plots of $\bar{V_x},\bar{V_y},V_x^{rms},V_y^{rms}$ versus $Re$ for four water-glycerin solutions both average velocity components and their rms fluctuations decrease smoothly with $Re$ from $2.5\times 10^5$ down to about $4000$ and the transition to turbulence is clearly observed at $Re\approx3300$ in the $\bar{V_y},V_y^{rms}$ dependencies on $Re$ in a good agreement with our measurements of $C_f$ for high viscosity solutions and early measurements published in Ref. \cite{ravelet}. But the dependencies of $\bar{V_x},V_x^{rms}$ on $Re$ do not show any sign of the transition at $Re\approx3300$. On the other hand, our measurements of rms fluctuations of velocity as a function of $Re$ contradict to the data presented in Ref. \cite{ravelet}, where a saturation of $V^{rms}$ at $Re\geq 3300$ was found.

\indent We further performed the experimental test of two possible sources of the violation of the similarity law, namely the non-stationary torque statistics (mentioned above), and the existence of other control parameters, besides $Re$. We demonstrated that at $\Omega> 3$ rad/s, the torque statistics is stationary, and a time series for about one hour is sufficient to characterize its average and rms fluctuations value. To test the presence of additional control parameters, such as $Pr$ and $Ra$, different pairs of water-glycerine solutions with the same viscosities in a pair but different $Pr$ and $Ra$ were used. Then the measurements of $\bar{\Gamma}$ versus $Re$ for both solutions in the pair could reveal whether these parameters play an essential role and provide different dependencies of $C_f$ on $Re$. Unfortunately, since these measurements should be performed at low $\Omega$, the torque measurements with the large torque meter do not have sufficient resolution and accuracy to reveal differences if any.

\indent Since the divergences of $C_f$ occurred at low rotation speeds, we conjectured that it arose due to limited device resolution. Thus we need to increase signal-to-noise ratio at low rotation speeds to better resolve the signal. To achieve this we used a higher resolution torque meter, which allowed us to better analyze the signal and find out the source of the noise. Moreover, errors with the small torque meter were also reduced, since a smaller torque meter required less bulky couplers that were difficult to align. A switch to smaller couplers inherently have to be better aligned due to geometrical restrictions.

\subsubsection{Small {range} torque meter}

\indent To further increase the resolution of the torque measurements and discover the source of irreproducibility in $\Gamma_{err}$ a much more sensitive and higher resolution torque meter was used in the same experimental setup. Installing this newer, physically smaller and lighter torque meter lowered the offset and irreproducibility of the measurements, but did not eliminate them. Instead of $0.3$ Nm the offset was reduced to $5$ mNm and irreproducibility brought down to a level of 10 mNm (inset of Fig. \ref{fig:CfTvsReprecise2disks}), but now thirty times larger than the torque meter resolution. Such drastic reduction allowed us to reliably measure smaller torques, but still we could not simply subtract $\Gamma_{err}$ from the average torque to remove the offset which in principle is still not the correct procedure.  Figure \ref{fig:TvsOmegaprecise2disks} shows the dependence of $\bar{\Gamma}$ corrected by a shift of $5$ mNm in a wide range of water-glycerin solution viscosities and $\Omega$ starting from $\sim1.2$ rad/s (or $f\simeq11.46$ rpm). Due to smaller range of the torque meter, only glycerol-water mixtures with viscosity up to  $18.86~mPa\cdot s$ were used in these experiments. The data are smooth with very small scatter even on the plot with larger resolution shown in the inset in Fig. \ref{fig:TvsOmegaprecise2disks}. The resulting friction coefficient {is constant in a wide range of $Re$ values from about $2\times 10^5$ down to $Re_c\approx3\times 10^3$, where $Re_c$ is the transition to turbulence (see Fig. \ref{fig:CfTvsReprecise2disks}a), though the constant value of $C_f$ is smaller due to new low friction ball-bearings and seals used. The dependence of $\Gamma_{rms}/\bar{\Gamma}$ on $Re$ is shown in Fig. \ref{fig:CfTvsReprecise2disks}b.

\indent The torque resolution is sufficient to compare different pairs of water-glycerine solutions with the same viscosities in a pair but different $Pr$ and $Ra$. The comparison for the first pair of water at $10.5^{\circ}C$ and a water-glycerin solution of $30\%$ w/w glycerin at $44^{\circ}C$ with the same $\eta=1.3~mPa\cdot s$, $Pr_w=9.63$ and $Pr_{Gl30}=9.53$, respectively, but $Ra_w/Ra_{Gl30}\approx7$ revealed a significant difference in $C_f$ at smaller values of $Re$ (see Fig. \ref{fig:CfvsRecomp2solutions}a). Contrarily, the comparison for another pair of water-glycerin solutions $10\%$ w/w glycerin at $10^{\circ}C$ and $40\%$ w/w glycerin at $45^{\circ}C$ with $\eta=1.74~mPa\cdot s$ and $1.75~mPa\cdot s$, respectively, almost the same $Pr\simeq12.3$ but different $Ra$ with ratio $Re_{Gl40}/Re_{GL10}\approx4$, does not show any significant difference (see Fig. \ref{fig:CfvsRecomp2solutions}b). Since we used  a zero frequency offset shift in $\bar{\Gamma}_0$ to compute $C_f$  instead of subtracting irreproducible and noisy $\Gamma_{err}$, which is the likely source of the contradicting results for two pairs of solutions,  the significance of $Ra$ as an additional parameter is doubtful and unlikely and we claim that second possible source of the validation of the similarity law gives a negative result: there is only one control parameter in the flow, namely $Re$.

\indent  Even though we significantly reduced $\Gamma_{err}$, it remained irreproducible and about 30 times larger than the torque meter resolution. We obtained $Re$ independent friction factor in the wide range of $Re$ seemingly resolving the breaking of self-similarity puzzle but using a shift to zero of $\bar{\Gamma}_0$, an unfounded approach, instead of subtracting $\Gamma_{err}$.  The sensitive torque meter allowed us a separate insight; the presence of slight misalignments of the torque meter rotation axis with the upper disk and motor shaft and a separate misalignment of the lower motor shaft can nonlinearly couple to produce a dc offset in the torque signal. The irreproducibility arose from minute mechanical movements of the independent parts of the apparatus due to the swirling of a large body of fluid, that changed the $"phases"$ of the individual components relative to one another and changed the dc offset. Since it was impossible to sufficiently precise align the upper and lower drives due to the existing design, we switched to a one-disk arrangement in the second setup.

\subsection{ Torque error estimates and experimental details}

\indent As we concluded above, the observed anomalous divergence of $C_f$ from its constant value at $Re$, which depends solely on $\eta$, is explained by the presence of irreproducible and noisy $\Gamma_{err}$, which consists of both frictional losses and misalignment errors, though irreproducibility results mostly from misalignments. To understand the cause of this observed divergence and divulge the unknown error mechanism, we switched to a more sensitive torque meter with $0.35$ Nm range with a precision of $0.1\% $ or $0.35 ~mNm$, enough to resolve the small torques at low rotation speeds. However, even with these sensitive torque measurements the divergences in $C_f$ persisted. In this case, the total measured torque $\bar{\Gamma}_0$ includes two parts: the signal $\Gamma_{sig}\equiv \bar{\Gamma}$ and the background errors $\Gamma_{err}$, which consist of friction from the ball-bearings and seals and misalignments. Then the total friction coefficient is $C_f=(\bar{\Gamma}+\Gamma_{err})/\rho\Omega^2 R^5$. The expression for $C_f$ can be rewritten via $Re$ as $C_f=(\bar{\Gamma}+\Gamma_{err})\rho/{Re}^2\eta^2 R$. In the fully turbulent state $C_f$ is a constant independent of $Re$ as shown above and in Refs. \cite{ravelet,burnishev}. From the expression for $C_f$ one finds that, if $\Gamma_{err}$>>$\bar{\Gamma}$, then one gets $C_f\sim 1/Re^2$. Hence the $Re$, at which the background errors become significant, is given by the expression $Re_t\approx\sqrt{\Gamma_{err}\rho/C\eta^2 R}$, where $C$ is the constant value of $C_f$ in the turbulent regime.

\indent Substituting $\Gamma_{err}$ with the value of the measured background, for example, in the case of the small precise torque meter discussed at the end of the previous subsection, one gets a value of $Re_t$, at which $C_f$ deviates from the constant value. Then for $\Gamma_{err}\approx 10$ mNm (see the inset in Fig. \ref{fig:CfTvsReprecise2disks}a) one gets for water at $T=10^{\circ}C$ (see Table 3)  $Re_t=3.1\times 10^4$ that agrees rather well with the lowest $Re$ observed for water in the experiment presented in Fig. \ref{fig:CfTvsReprecise2disks}a. However, the deviation in $C_f$ from the constant value still does not show up there, since $Re_t$ corresponds to $\Omega\simeq 2$ rad/s, below which the measurements were not extended due to larger scatter. Similar accordance is found for the rest of data shown in Fig. \ref{fig:CfTvsReprecise2disks}a, and at each $\eta$ the measurements were ceased for $\Omega$ between 1.2 and 2 rad/s, where the scatter is enhanced and $C_f$ diverges. On the other hand, in the case of measurements with the large torque meter, $\Gamma_{err}$ was an order of magnitude or more larger and divergence of $C_f$ from the constant value occurred at $\Omega> 3$ rad/s, at which the measurements were still conducted. For example, for water at $T=15^{\circ}$C we obtain $Re_t\simeq9.5\times10^4$ a factor 3-4 smaller than the value of $Re$ observed in the experiment, at which $C_f$ starts to deviate from the constant (see Fig. \ref{fig:CfvsRelarge}). The value of $Re_t$ corresponds to $\Omega\simeq 5.5$ rad/s,  at which $\bar{\Gamma}_0\simeq 0.16$ Nm and $C_f$ exceeds the constant by $\sim15\%$. It means that the real value of $\Gamma_{err}$ is  $15\%$ of the signal $\bar{\Gamma}$, and its contribution to $C_f$ grows with a decreasing rotation speed as $\Omega^{-2}$. Thus the divergence measured with the large torque meter can be attributed to either an unjustified shift to zero of the measured $\bar{\Gamma}_0$ or an incorrect subtraction of the background signal, which is stable and approximately constant for high rotation {speeds} but noisy and irreproducible at low rotation speeds.

\subsection{Precise measurements of torque in a swirling flow driven by a single disk}

\indent In spite of a reduction in $\Gamma_{err}$  achieved in the one-disk configuration (see inset in Fig. \ref{fig:Terronedisk}) with careful measurements using the small precise torque meter, they were still a factor of 20 larger than the resolution of the device, $0.35$ mNm. The measured background was also still irreproducible and noisy preventing correct calibration of the device for $\Omega <10~ rad/s$. Since a large volume of a working fluid is driven by a large and powerful motor, it is not surprising that slight misalignments between the motor, torque meter and disk axes strongly affects the torque measurements. Due to the misalignment the shaft attached to the spinning disk undergoes a slight nutation which is exaggerated by the height of couplers used to attach to the torque meter. The misalignment of the torque meter rotation axis and the motor shaft and the nutation of the disk shaft coupler causes small periodic variations in the measured torque, which if occurring singly can be filtered out from the measured signal. But both these error mechanisms exist and in unison couple nonlinearly to produce a DC offset in the measured torque and cannot be filtered out. Hence the background signal measured is a sum of the friction in the bearings and seals that is small {in comparison} and the DC offset caused by the misalignment, and $\Gamma_{err}$ consists of terms due to both the frictional losses and misalignment errors.

\indent In order to perform precise torque measurements at low frequencies and extend the range of the torque meter, the background errors were minimized by an alignment of the torque meter rotation axis with the disk and motor shafts, as it described in Section II. First, time series of the torque measurements at a constant $\Omega$ was used to test the stability of the measurements from the deviation of $\bar{\Gamma}$ from its constant value. The degree of long-term-stability of $\bar{\Gamma}_0$ for the glycerin-water solutions with $c=50\%$ w/w of glycerin at $20.2 ~^{\circ}$C  and $Re=10^4$ and $c=85\%$ w/w of glycerin at $20.2 ~^{\circ}$C and $Re=10^3$ can be estimated from the data presented in Fig. 5SM in \cite{sm}, where only a part of the data is shown. Stationarity of $\Gamma$ statistics is demonstrated by Gaussian probability distribution functions in Fig. 6SM and flat torque power spectra in Fig. 7SM in \cite{sm} for both solutions.

\indent After careful alignment background errors  were reduced to $\sim 3$ mNm at $\Omega>6$ rad/s, about 8.5 times larger than the device resolution and almost a factor 30 smaller than that measured with the large torque meter (see Fig. \ref{fig:Terronedisk}). The value of the background increases at $\Omega <6$ rad/s up to 2 rad/s by the factor 1.6 but in the whole range of $\Omega$ it is reproducible for different experimental runs and shows small scatter Fig. \ref{fig:Terronedisk}.  After these modifications the background DC signal was stable and reproducible to within 0.2 mNm close to the precision allowed by the instrument. Thus after the correct background subtraction and filtering the torque signal first by a 2Hz low pass filter to remove the PID signal and second by a notch filter to remove the drive frequency, the measured torque is precise with 1 mNm due to any misalignment errors and free from any divergence (Fig. \ref{fig:TvsOmegabladed}). Using similar analysis as applied above, the divergence of $C_f$ for water at $T=10^{\circ}$C is calculated to be at $Re\sim10^4$, which is now below our parameter range. However even after the careful alignment and measurement of the background torque, small deviations from the scaling dependence were observed for low rotation speeds. This stems from the fact that the background signal, specifically the nonlinear interaction between the disk shaft and motor shaft is modified in presence of fluid (instead of air) that is impossible to estimate; for this reason measurements below $1.5-2$ rad/s depending on $\eta$ were unreliable and discarded (see Fig. \ref{fig:CfTrmsTvsRe1disk}). By reducing {alignment errors} we, first, were  able to cut $\Gamma_{err}$ down to the resolution of the torque meter, and, second, to get the value of $Re$ at which the deviation from the self-similarity takes place in an agreement with the estimates. The dependence of $\Gamma_{rms}/\bar{\Gamma}$ on $Re$ is presented in the inset in Fig. \ref{fig:CfTrmsTvsRe1disk}.

\section{Conclusion}

\indent Scrupulous measurements and detailed data analysis of the torque in a swirling turbulent flow driven by the counter-rotating bladed disks revealed an apparent breaking of the law of similarity that could arise from several possible factors including dependence on dimensional numbers other that $Re$ or velocity coupling to other fields such as temperature. However careful redesign and calibration of the experiment showed this unexpected result was due to background errors caused by  minute misalignments which lead to a noisy and irreproducible torque signal at low rotation speeds and prevented correct background subtraction normally ascribed to frictional losses.  An important lesson to be learnt is that multiple minute misalignments can nonlinearly couple to the torque signal and provide a dc offset that cannot be removed by averaging and cause the observed divergence of the friction coefficient $C_f$ from its constant value observed in the turbulent regime. To minimize the friction and misalignments, we significantly modified the experimental setup and carried out the experiment with one bladed disk where the disk, torque meter and motor shaft axes can be aligned with significantly smaller error, close to the torque meter resolution. As a result we made precise measurements with high resolution and sensitivity of the small torques produced for low rotation speeds for several water-glycerin solutions of different viscosities and confirmed the similarity law in a  wide range of $Re$ in particular in low viscosity fluids.

\section*{Disclosure statement}

No potential conflict of interest was reported by the authors

\section*{Acknowledgements}

We thank Guy Han for his excellent technical support. 

\section*{Funding}

This work was partially supported by the Israel Science Foundation (ISF; grant \# 882/15).

\begin{figure}
\centering
\includegraphics[width=9cm]{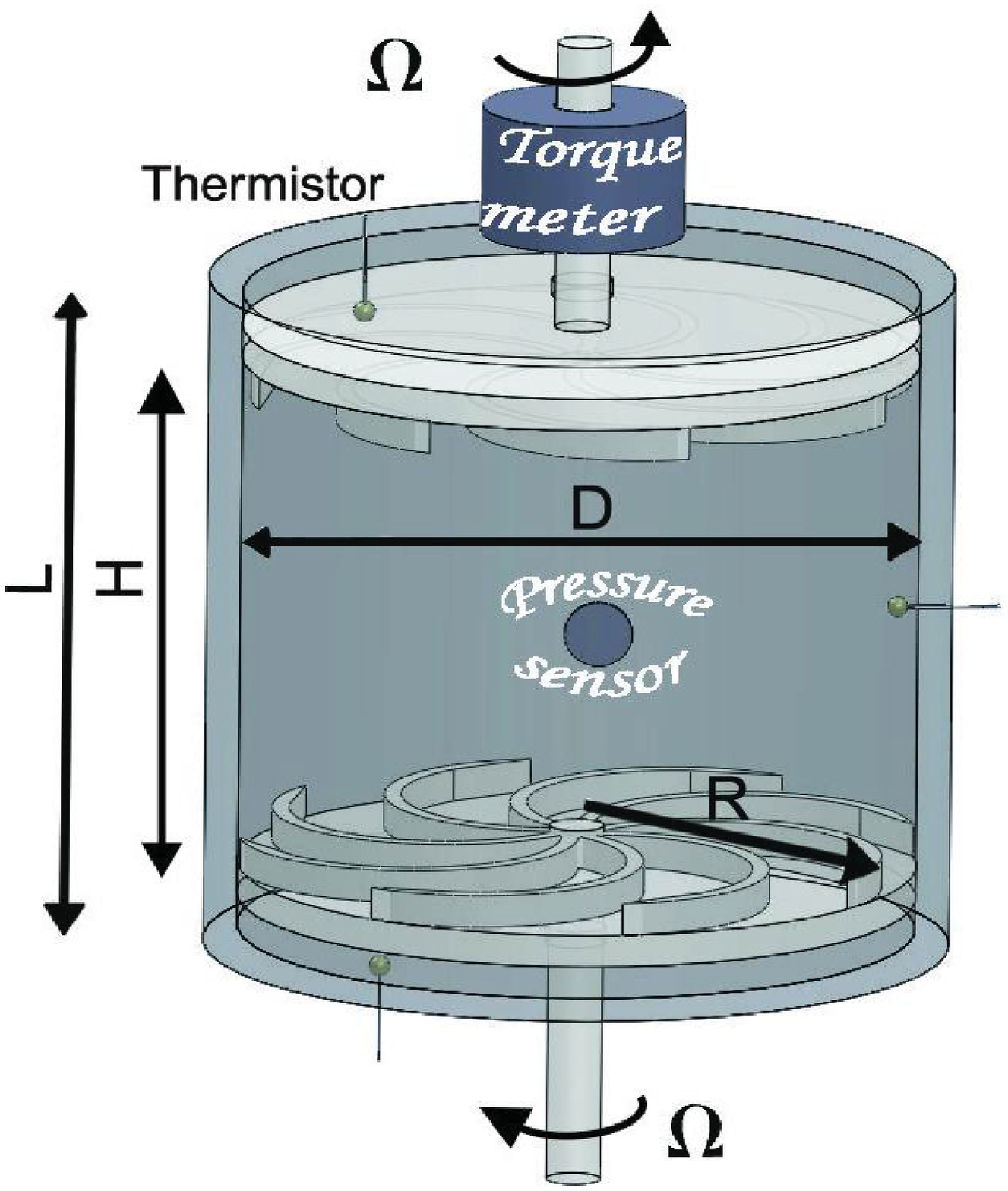}

\caption{(Color online) Experimental setup with two counter-rotating bladed disks}

\label{fig:setup1}

\end{figure}

\begin{figure}
\includegraphics[width=17cm]{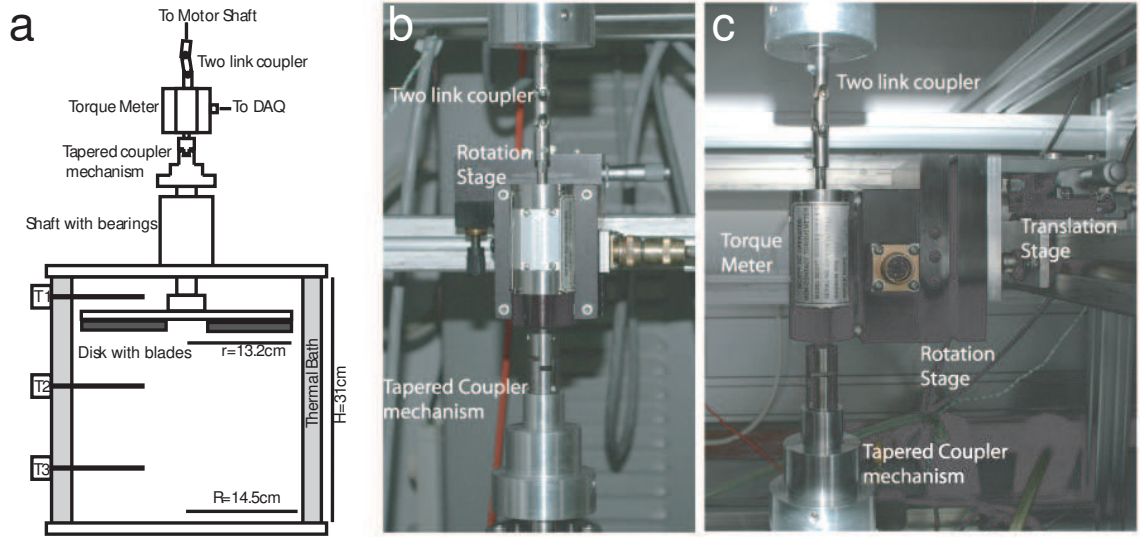}

\caption{Experimental setup with one rotating bladed disk: (a) Schematic experimental setup with details of couplers; (b) Front view of two translational and two rotation stages; (c) Side view of two translational and two rotation stages.}

\label{fig:setup2}

\end{figure}

\begin{figure}
\includegraphics[width=11cm]{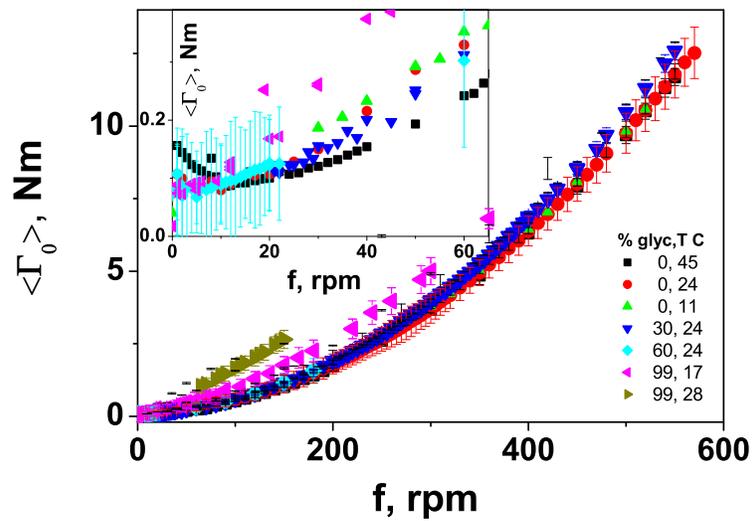}

\caption{(Color online) The average torque $\langle\Gamma_0\rangle$ as a function of rotation frequency $f$ in a wide range of its variations down to 1 rpm for water-glycerin solutions in a wide range of w/w glycerin concentrations from $0\%$ till $99\% $ and various temperatures for the first setup with two counter-rotating bladed disks. Inset: the same data at higher resolution at low $f$. }

\label{fig:Tvsomegaoldfull}
\end{figure}

\begin{figure}
\includegraphics[width=11cm]{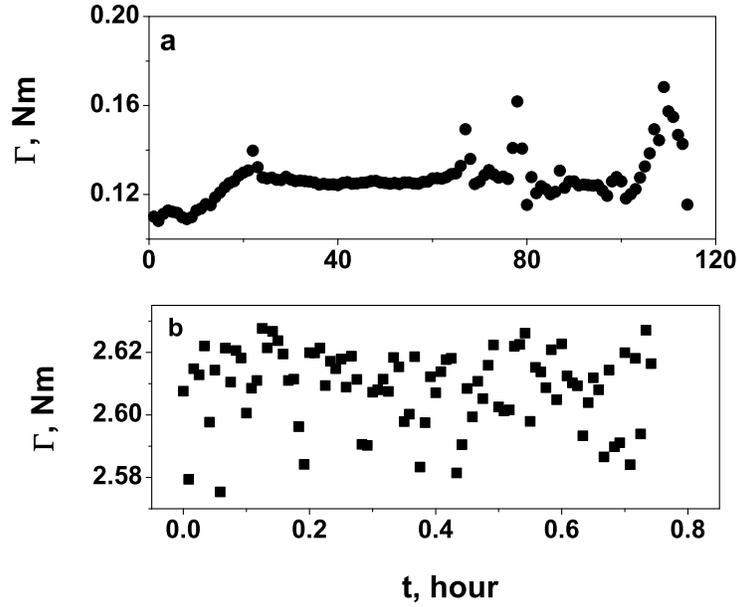}

\caption{(a) Very long time series of torque $\Gamma(t)$ after switching from $f=13$ rpm and $T=42^{\circ}$ to $f=26$ rpm and $T=24^{\circ}$ for water-glycerin solution with $60\%$ w/w glycerin. (b) Time series of $\Gamma(t)$ and temperature $T$ at $f=240$ rpm ($Re=1.1\times 10^5$) for water-glycerin solution with $20\%$ w/w glycerine at $T\simeq21^{\circ}$. }

\label{fig:Tvst60Gl24C}
\end{figure}

\begin{figure}
\includegraphics[width=9cm]{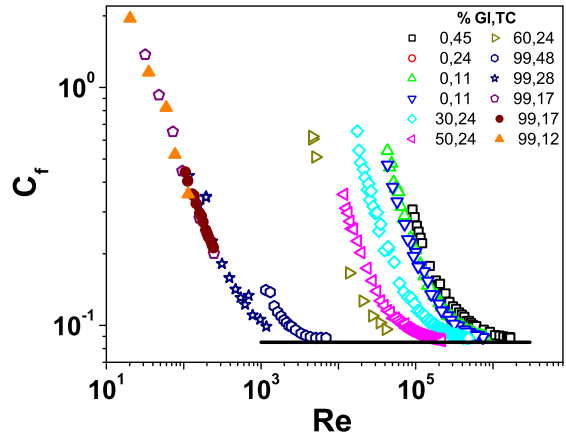}

\caption{(Color online) First run for the friction coefficient $C_f$ versus $Re$ for water-glycerine solutions in a wide range of w/w glycerin concentrations from $0\%$ till $99\% $ and various temperatures for the first setup. Solid line is to indicate a constant value of $C_f\simeq0.085$ in a fully developed turbulent state.}

\label{fig:CfvsReold}

\end{figure}

\begin{figure}
\includegraphics[width=11cm]{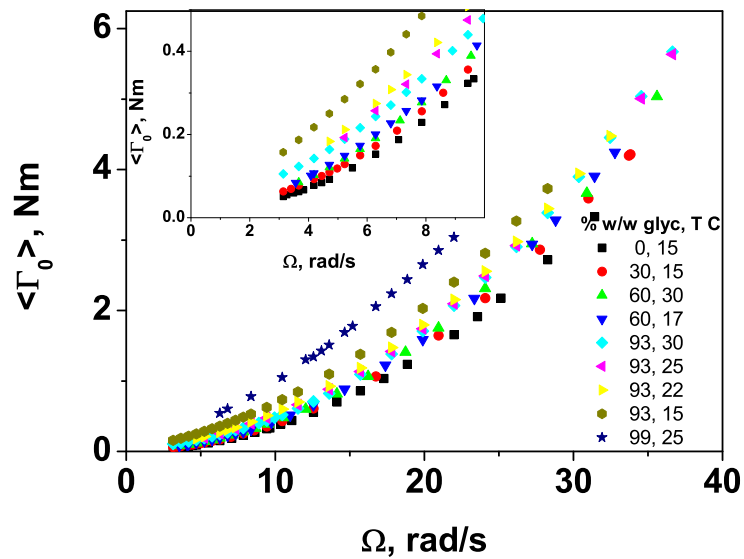}

\caption{(Color online) The average torque $\langle\Gamma_0\rangle$ as a function of angular velocity $\Omega$ in a wide range of its variations down to 3 rad/s (or about 30 rpm) for water-glycerin solutions in a wide range of w/w glycerin concentrations from $0\%$ till $99\% $ and various temperatures for the first setup.Inset: the same data at higher resolution at low $f$.}

\label{fig:TvsOmegalarge}

\end{figure}

\begin{figure}
\includegraphics[width=11cm]{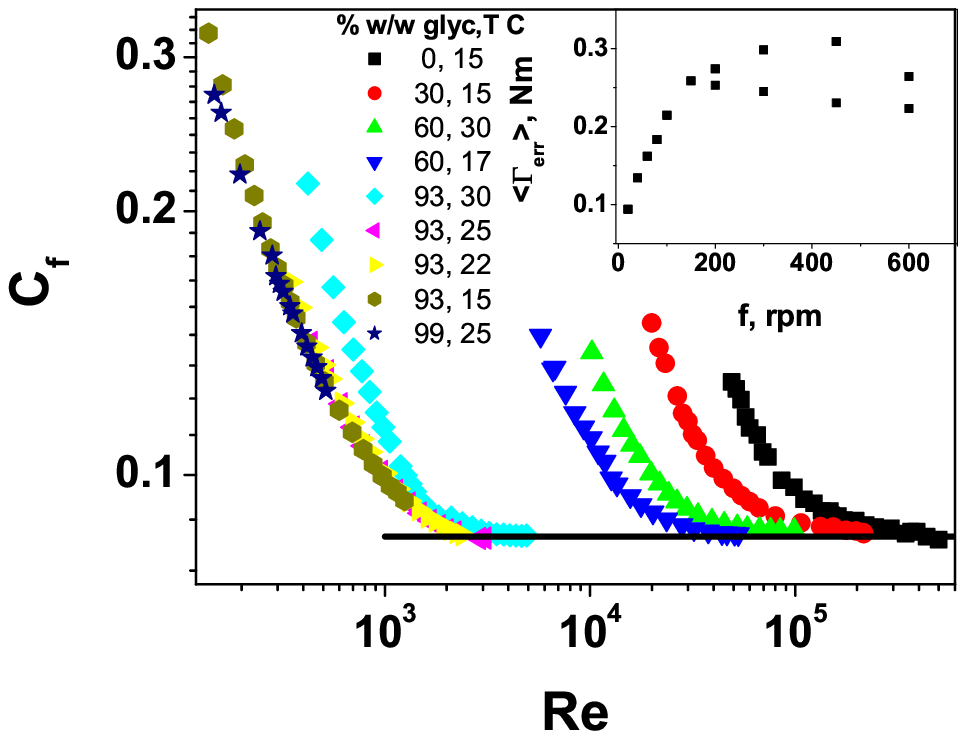}

\caption{(Color online) Second run for the friction coefficient $C_f$ versus $Re$ for water-glycerine solutions in a wide range of w/w glycerin concentrations from $0\%$ till $99\% $ and various temperatures for the first setup. Solid line is to indicate a constant value of $C_f\simeq0.085$ in a fully developed turbulent state.}

\label{fig:CfvsRelarge}

\end{figure}

\begin{figure}
\includegraphics[width=11cm]{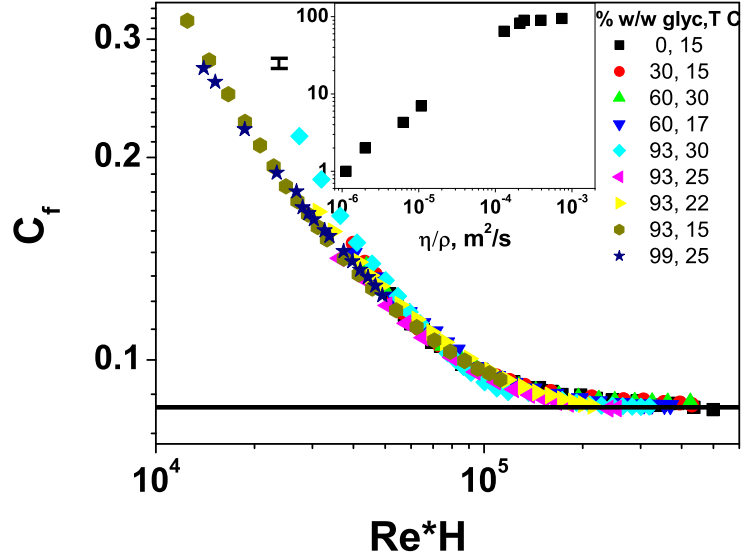}

\caption{(Color online) The same data as shown in Fig. \ref{fig:CfvsRelarge} with $Re$ scaled by $H(f)$ a function of viscosity only shown in the Inset, which is chosen to collapse the data for different viscosities. olid line is to indicate a constant value of $C_f\simeq0.085$ in a fully developed turbulent state.}

\label{fig:CfvsReHlarge}

\end{figure}

\begin{figure}
\includegraphics[width=10cm]{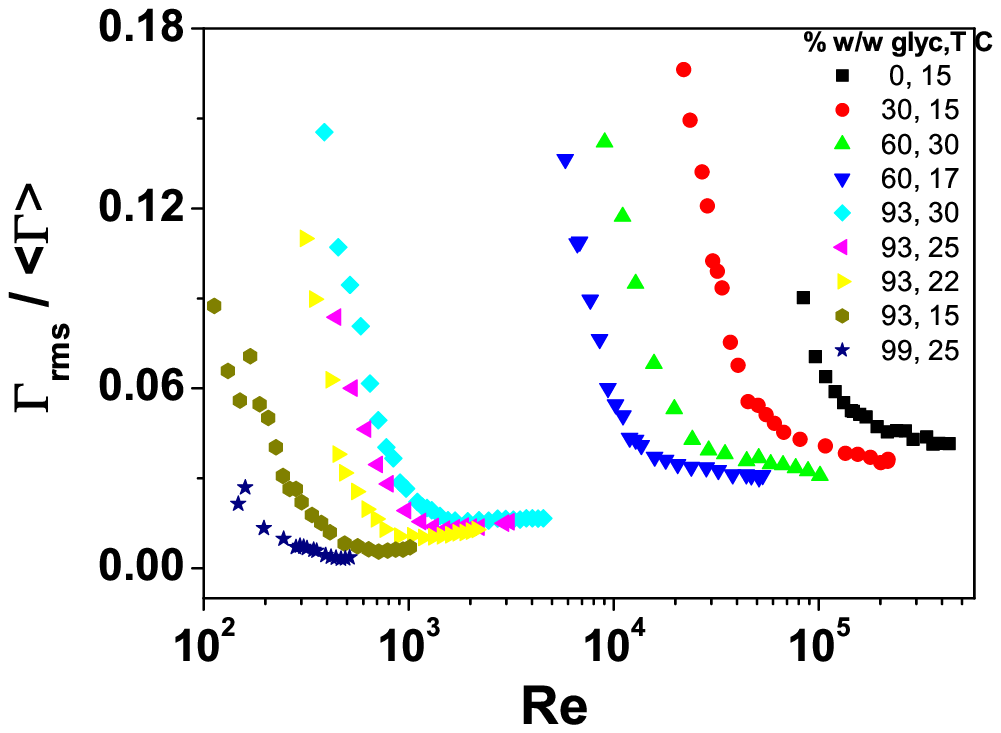}

\caption{(Color online) Turbulent intensity $\Gamma_{rms}/\langle\Gamma\rangle$ versus $Re$ for water-glycerine solutions in a wide range of w/w glycerin concentrations from $0\%$ till $99\% $ and various temperatures for the first setup. }

\label{fig:TrmsTvsRelarge}
\end{figure}

\begin{figure}
\includegraphics[width=9cm]{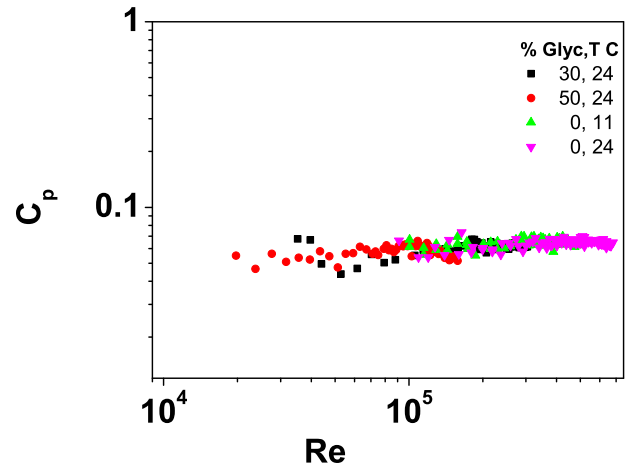}

\caption{(Color online) Normalized pressure fluctuations $C_p$ versus $Re$ for three water-glycerine solutions and various temperatures.}

\label{fig:CpvsReold}
\end{figure}

\begin{figure}
\includegraphics[width=15cm]{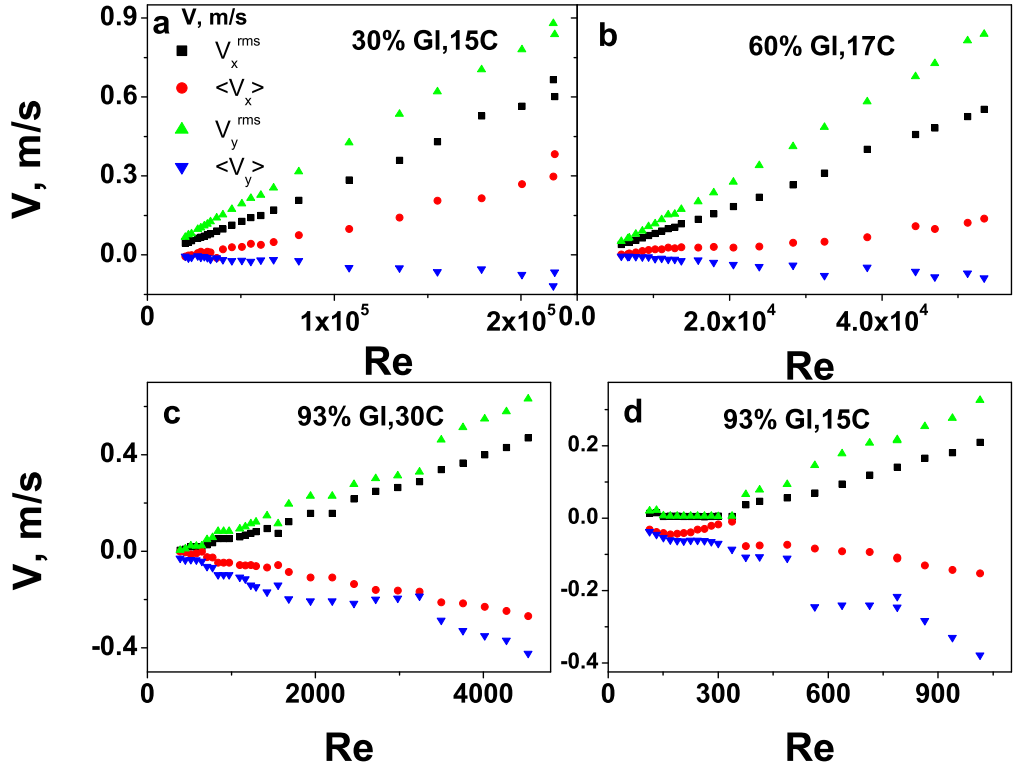}

\caption{(Color online) Dependence of two components of average $\langle V_x\rangle$, $\langle V_y\rangle$ and rms fluctuation $V_x^{rms}$, $V_y^{rms}$ velocities on $Re$ for 4 water-glycerin solutions for the first setup:(a) $30\%$ w/w glycerin at $15~^{\circ}C$; (b) $60\%$ w/w glycerin at $17~^{\circ}C$; (c) $93\%$ w/w glycerin at $30~^{\circ}C$; (d) $93\%$ w/w glycerin at $15~^{\circ}C$. }

\label{fig:VxVyvsRe}
\end{figure}

\begin{figure}
\includegraphics[width=11cm]{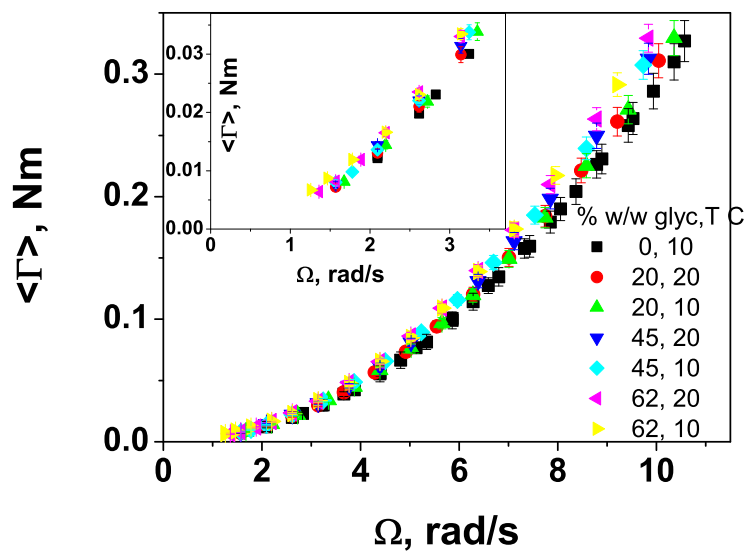}

\caption{(Color online) The average torque $\langle\Gamma\rangle$ as a function of angular velocity $\Omega$ in a wide range of its variations down to $\simeq 1.2$ rad/s for water-glycerin solutions in a wide range of w/w glycerin concentrations and various temperatures for the first setup but with a high precision torque meter. Inset: the same data at higher resolution at low $\Omega$. }

\label{fig:TvsOmegaprecise2disks}
\end{figure}

\begin{figure}
\includegraphics[width=11cm]{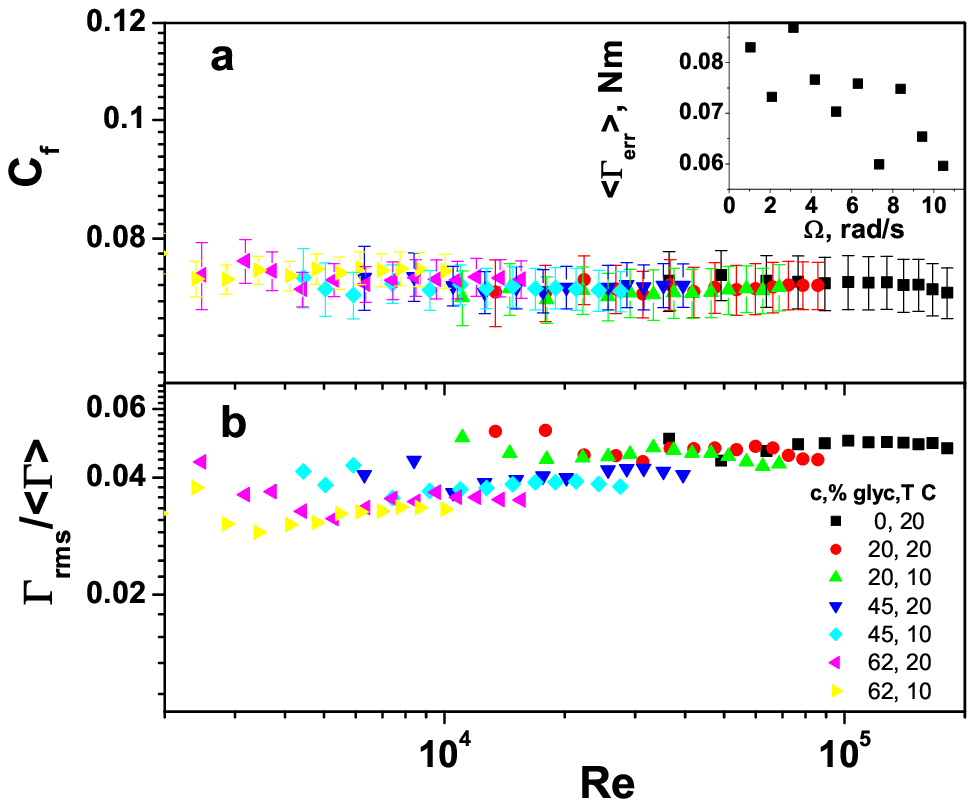}

\caption{(Color online) (a) Friction coefficient $C_f$ versus $Re$ for water-glycerine solutions in a wide range of w/w glycerin concentrations and various temperatures for the first setup but with a high precision torque meter. Inset: A single realization of the average torque background error $\langle\Gamma_{err}\rangle$ values versus $\Omega$. (b) Turbulent intensity $\Gamma_{rms}/\langle\Gamma\rangle$ versus $Re$ for water-glycerine solutions at the same conditions.}

\label{fig:CfTvsReprecise2disks}
\end{figure}

\begin{figure}
\includegraphics[width=9cm]{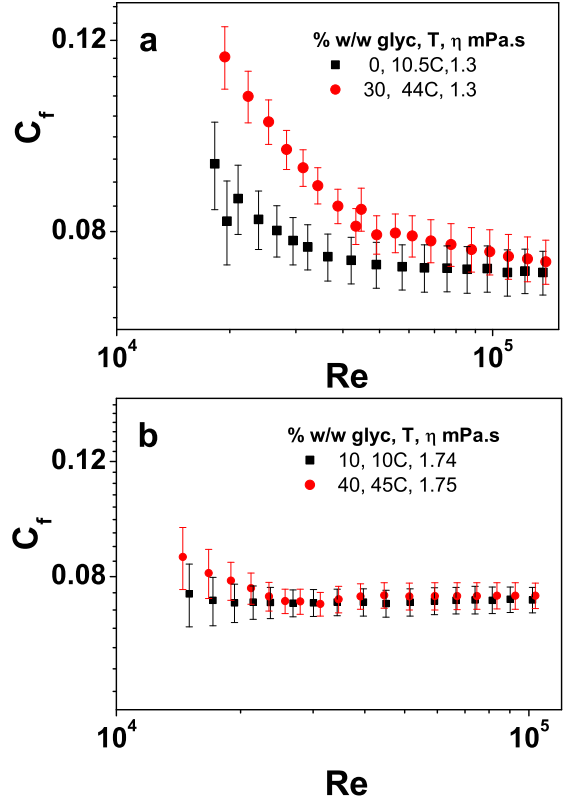}

\caption{(Color online) (a) Friction coefficient $C_f$ versus $Re$ for two water-glycerine solutions with the same $\eta=1.3~mPa\cdot s$ and of $0\%$ w/w glycerin concentrations and $T=10.5^{\circ}C$ and $30\%$ w/w glycerin concentrations and $T=44^{\circ}C$ for the first setup but with a high precision torque meter. (b) $C_f$ versus $Re$ for another two water-glycerine solutions with the same $\eta\simeq1.75~mPa\cdot s$ and of $10\%$ w/w glycerin concentrations and $T=10^{\circ}C$ and $40\%$ w/w glycerin concentrations and $T=45^{\circ}C$ in the same setup. }

\label{fig:CfvsRecomp2solutions}
\end{figure}

\begin{figure}
\includegraphics[width=9cm]{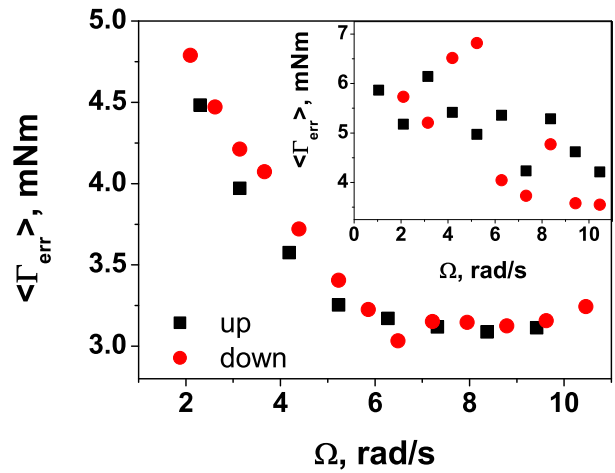}

\caption{(Color online) Two realizations (back and forth) of the average torque background error $\langle\Gamma_{err}\rangle$ values versus $\Omega$ for the second setup with a single rotating bladed disk and a high precision torque meter. Inset: Two realizations (back and forth) of the average torque background error $\langle\Gamma_{err}\rangle$ values versus $\Omega$ in the same setup.  }

\label{fig:Terronedisk}
\end{figure}

\begin{figure}
\includegraphics[width=11cm]{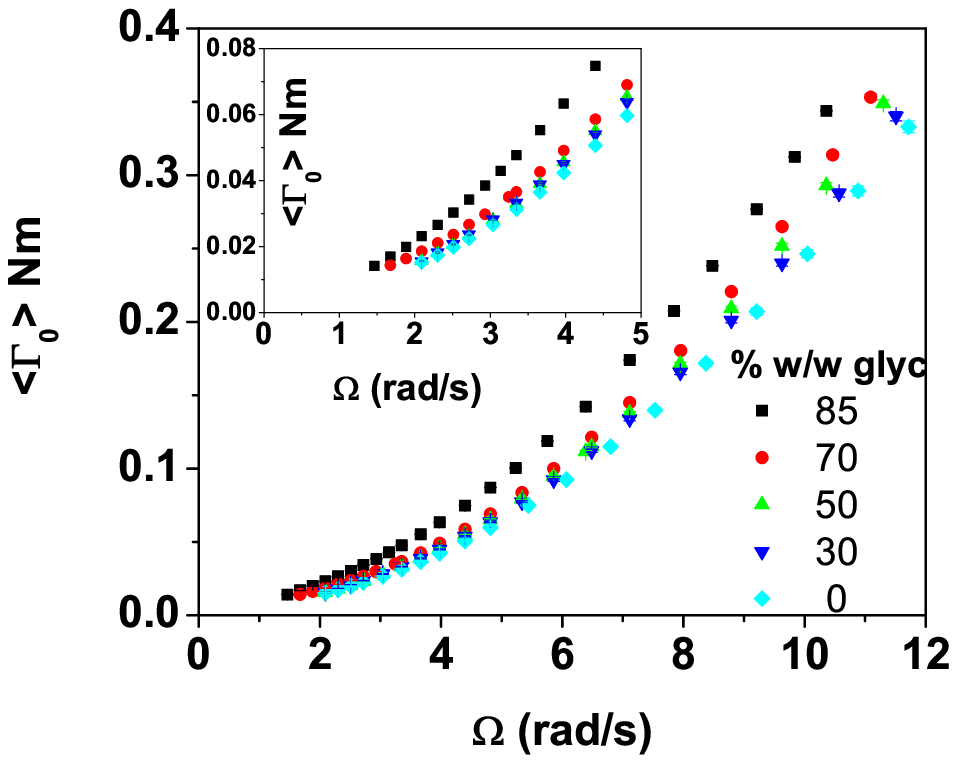}

\caption{(Color online) The average torque $\langle\Gamma_0\rangle$ as a function of angular velocity $\Omega$ in a wide range of its variations down to $\simeq 1.5$ rad/s for water-glycerin solutions in a wide range of w/w glycerin concentrations and temperature $T=24^{\circ}C$ for the second setup with a single rotating bladed disk and a high precision torque meter. Inset: the same data at higher resolution at low $\Omega$. }

\label{fig:TvsOmegabladed}
\end{figure}

\begin{figure}
\includegraphics[width=10cm]{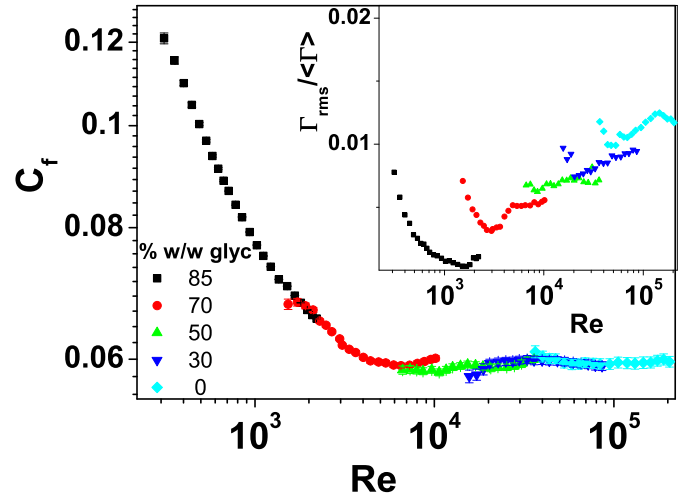}

\caption{(Color online) Friction coefficient $C_f$ versus $Re$ for water-glycerine solutions in a wide range of w/w glycerin concentrations and temperature $T=24^{\circ}C$ for the second setup with a single rotating bladed disk and a high precision torque meter. Inset: Turbulent intensity $\Gamma_{rms}/\langle\Gamma\rangle$ versus $Re$ for the same water-glycerine solutions and in the same setup. }

\label{fig:CfTrmsTvsRe1disk}
\end{figure}

\begin{table}{Table1}

\begin{center}
\begin{tabular}{|c|c|c|c|c|c|c|c|c|}
  \hline
  $c$ ($\%$ glycerine) &0&0&0&30&50&60&99&99\\
  \hline
  $T$($^{\circ}C$) &45&24&11&24&24&24&28&17\\
  $\eta$ ($mPa\cdot s$) &0.6&0.91&1.27&2.217&5.2&9.144&540&1432 \\
  \hline

\label{Table1}
\end{tabular}
\end{center}
\end{table}

\begin{table}[h]{Table2}
\begin{center}
\begin{tabular}{|c|c|c|c|c|c|c|c|c|c|}
  \hline
  $c$ ($\%$) &0&30&60&60&93&93&93&93&99\\
  \hline
  $T$($^{\circ}C$) &15&15&17&30&30&25&22&15&25\\
  \hline
  $\eta$ ($mPa\cdot s$) &1.12&3.04&7.2&12.8&172&269.5&328&613 &825 \\
  \hline
\label{Table2}
\end{tabular}
\end{center}
\end{table}

\begin{table}[h]{Table3}
\begin{center}
\begin{tabular}{|c|c|c|c|c|c|c|c|}
  \hline
  $c$ ($\%$) &0&20&20&45&45&62&62\\
  \hline
  $T$($^{\circ}C$) &10&20&10&20&10&20&10\\
  \hline
  $\eta$ ($mPa\cdot s$)
  &1.31&2.14&2.76&4.77&6.89&11.86&18.86 \\
  \hline
\label{Table3}
\end{tabular}
\end{center}
\end{table}

\begin{table}[h]{Table4}
\begin{center}
\begin{tabular}{|c|c|c|c|c|c|}
  \hline
  $c$ ($\%$) &0&30&50&70&85\\
  \hline
  $T$($^{\circ}C$) &20.0&20.2&20.2&20.0&20.2\\
  \hline
  $\eta$ ($mPa\cdot s$) &1&2.5&6.17&22.5&101.9 \\
  \hline
\label{Table4}
\end{tabular}

\end{center}
\end{table}


\begin{thebibliography}{40}
\bibitem{fauve} {Fauve S, Laroche C,  Castaing B,  Pressure fluctuations in swirling turbulent flows, J. Phys. II 1993;3:271.}
\bibitem{tabeling}{Zocchi G, Tabeling P, Maurer J, Williams H,  Measurement of the scaling of the dissipation at high Reynolds numbers, Phys. Rev. E 1994;50:3693.}
\bibitem{cadot}{Cadot O, Douady S, Couder Y, Characterization of the low-pressure filaments in a three-dimensional turbulent shear flow, Phys. Fluids 1995;7:630. }
\bibitem{labbe}{Labbe R, Pinton J-F, Fauve S, Power fluctuations in turbulent swirling flows, J. Phys. II 1996;6:1099 .}
\bibitem{pinton}{Mordant N, Pinton J-F, Chilla F, Characterization of turbulence in a closed flow, J. Phys. II 1997;7:1729.}
\bibitem{ravelet}{Ravelet F, Chiffaudel A, Daviaud F, Supercritical transition to turbulence in an inertially driven von Karman closed flow, J. Fluid Mech. 2008;601:339.}
\bibitem{burnishev}{Burnishev Yu, Steinberg V, Torque and pressure fluctuations in turbulent von Karman swirling flow between two counter-rotating disks. I, Phys. Fluids 2014;26:055102.}
\bibitem{burnishev1}{Burnishev Yu, Steinberg V, Turbulence and turbulent drag reduction in swirling flow: Inertial versus viscous forcing, Phys. Rev. E 2015;92:023001.}
\bibitem{unpublished_data} {Burnishev Yu, Steinberg V,  unpublished data. 2014.}
\bibitem{Dar} {Falkovich G, Darrigol K.O, Worlds of Flow. Oxford Univ. Press, 2005).}
\bibitem{batchelor}{ Batchelor G. K., An Introduction to Fluid Dynamics. Cambridge Univ. Press, 1967.}
\bibitem{landau} {Landau L, Lifshitz E, Fluid Mechanics. Pergamon, London, 1959.}
\bibitem{falkovich} {Falkovich G, Fluid Mechanics, a short course for physicists. Cambridge Univ. Press,  2011.}

\bibitem{aryesh}{Mukherjee A, Steinberg V, Von Karman swirling flow between a rotating and a stationary smooth disk: Experiment, 2017.}
\bibitem{sm}{Supplemental Material}


\end{thebibliography}
\end{document}